\documentclass[a4paper,11pt]{article}

\usepackage{amsthm}
\usepackage{amsmath}
\usepackage{amssymb}
\usepackage{amsfonts}
\usepackage{amstext}
\usepackage{xspace}
\usepackage{graphicx}
\usepackage{mathrsfs}
\usepackage{graphics}

\usepackage{algorithm}
\usepackage{algorithmic}
\usepackage{fullpage}

\title{Polynomial Time Efficient Construction Heuristics for Vertex Separation Minimization Problem}

\author{Pallavi Jain\thanks{Department of Mathematics, Dayalbagh Educational Insitute, Agra, India \texttt{pallavijain.t.cms@gmail.com}}
\and Gur Saran\thanks{Department of Mathematics, Dayalbagh Educational Insitute, Agra, India \texttt{gursaran@dei.ac.in}}
\and Kamal Srivastava\thanks{Department of Mathematics, Dayalbagh Educational Insitute, Agra, India \texttt{kamal.sri@dei.ac.in}}
}

\date{}

\begin{document}

\maketitle
% DO NOT REMOVE: Creates space for Elsevier logo, ScienceDirect logo
% and ENDM logo
%\begin{verbatim}\end{verbatim}\vspace{2.5cm}

%\begin{frontmatter}

\begin{abstract}
Vertex Separation Minimization Problem (VSMP) consists of finding a layout of a graph $G = (V, E)$ which minimizes the maximum vertex cut or separation of a layout. It is an NP-complete problem in general for which metaheuristic techniques can be applied to find near optimal solution. VSMP has applications in VLSI design, graph drawing and computer language compiler design. VSMP is polynomially solvable for grids, trees, permutation graphs and cographs. Construction heuristics play a very important role in the metaheuristic techniques as they are responsible for generating initial solutions which lead to fast convergence. In this paper, we have proposed three construction heuristics \emph{H}1, \emph{H}2 and \emph{H}3 and performed experiments on Grids, Small graphs, Trees and Harwell Boeing graphs, totaling 248 instances of graphs. Experiments reveal that \emph{H}1, \emph{H}2 and \emph{H}3 are able to achieve best results for 88.71\%, 43.5\% and 37.1\% of the total instances respectively while the best construction heuristic in the literature achieves the best solution for 39.9\%
of the total instances. We have also compared the results with the state-of-the-art metaheuristic GVNS and observed that the proposed construction heuristics improves the results for some of the input instances. It was found that GVNS obtained best results for 82.9\% instances of all input instances and the heuristic \emph{H}1 obtained best results for 82.3\% of all input instances.
\end{abstract}
%
%\begin{keyword}
%Vertex Separation, Construction Heuristics, Graph Layout.
%\end{keyword}

\section{Introduction}\label{intro}

\begin{figure}
  \centering
  \includegraphics[height=1.43in,width=3.54in]{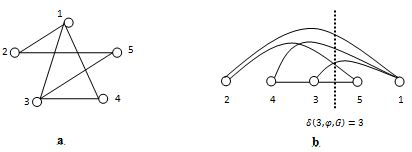}
  \caption{(a) A graph and (b) layout $\varphi=(2,4,3,5,1)$}\label{fig1}
\end{figure}

Graph layout problems are a class of combinatorial optimization problems whose goal is to find a layout of an input graph \emph{G} to optimize a certain objective function. A linear layout or layout of an undirected graph $G = (V,E)$, where $|V| = n$ is the bijective function $\varphi \colon V\rightarrow[n]=\{1,2,\dots,\emph{n}\}$. Set of all layouts is denoted by $\Phi(G)$. A Region in layout is defined as the area between two consecutive vertices in the layout. Vertex Separation minimization problem (VSMP) is to find a layout $\varphi^* \in \Phi(G)$ of a graph $G = (V, E)$ which minimizes the vertex separation (\emph{VS}) where $VS=\max_{i\in[|V|]}⁡\delta(i,\varphi,G)$ for a layout $\varphi$ where, $\delta(i,\varphi,G)=|{u\in L(i,\varphi,G)\colon \exists v\in R(i,\varphi,G)\wedge(u,v)\in E(G)}|$, $L(i,\varphi,G)=\{u\in V \colon \varphi(u)\leq i\}$ and $R(i,\varphi,G)=\{u \in V \colon \varphi(u)>i\}$ \cite{diaz,petit}. VSMP is NP-complete in general and has applications in VLSI design, graph drawing and computer language compiler design \cite{duarte}.\\\\
Further, in this paper, a vertex identifier corresponds to the vertices of graphs which are assumed to be the natural numbers 1,\dots,\emph{n}. For a vertex $u \in V(G)$, neighbourhood of \emph{u} in \emph{G}, $ N(u)=\{v\in V(G)\colon (u,v)\in E(G)\}$ and degree of \emph{u} in \emph{G}, $ d_G (u)=|{v\in V(G)\colon
(u,v)\in E(G)}|$.
Fig. \ref{fig1}(b) represents a linear layout of Fig. \ref{fig1}(a). %In Fig. \ref{fig1}(b), $L(1, \varphi ,G)={2}$, $R(1,\varphi,G)=\{4,3,5,1\}$. Therefore $\delta(1,\varphi,G)=1$ as there is only one vertex in \emph{L} which is adjacent to the vertices in \emph{R}. Similarly, $\delta(2,\varphi,G)=2$ as $L(2,\varphi,G)=\{2,4\}$, $R(2,\varphi,G)=\{3,5,1\}$ and both the vertices of \emph{L} are adjacent to the vertices of \emph{R}. In a similar manner $\delta(3,\varphi,G)=3$ and $\delta(4,\varphi,G)=3$. Therefore,
\emph{VS} for this layout $\varphi$ of \emph{G} is 3.
VSMP is polynomially solvable for grids, trees, permutation graphs and cographs \cite{diaz,petit}. This problem has been explored using metaheuristics namely, variable neighbourhood search by \cite{duarte,oro} in which they have proposed construction heuristics to generate an initial solution. Noberto et al. \cite{noberto} highlight the importance of construction heuristics in metaheuristics. They have proposed eight construction heuristics and have compared all the construction heuristics. In this paper, we have proposed three construction heuristics \emph{H}1, \emph{H}2, \emph{H}3. An experiment has been performed to compare all the construction heuristics on the test instances of Grids, Small graphs, Trees and Harwell Boeing graphs, totaling 248 instances of graphs. Experiments reveal that \emph{H}1, \emph{H}2 and \emph{H}3 are able to achieve best results for 87.9\%, 43.5\% and 37.1\% of the total instances respectively, while the best construction heuristic in the literature achieves the best solution for 39.9\% of the total instances. We have also compared the results of \emph{H}1 with the state-of-the-art metaheuristic general variable neighbourhood search (GVNS) \cite{oro} and observed that the proposed construction heuristics improves the results for some of the input instances. It was also observed that GVNS obtains best results for 82.9\% instances of all input instances while the heuristic \emph{H}1 obtained best results for 82.3\% instances of all input instances.
Further, in this paper Section 2 presents the construction heuristics proposed in this paper followed by computational experiments to show the efficiency of heuristics in Section 3.
\section{Construction Heuristics}
In this section, three construction heuristics have been proposed. These heuristics construct a layout by iteratively adding vertices to an initially empty layout. In each iteration, heuristic \emph{H}1 tries to place a vertex \emph{u} in the partial layout which minimizes its contribution as well as the contribution of vertices which are already placed in the partial layout. Motivation behind \emph{H}2 and \emph{H}3 is that if adjacent vertices are closely placed then they contribute to fewer regions. Further, in order to minimize the number of regions to which a vertex contributes, vertices of smaller degree are placed first in the partial layout. In the following sections, we describe the three construction heuristics which are essentially greedy methods.
\subsection{Heuristic \emph{H}1}
This procedure is outlined in Algorithm 1. Let \emph{layout} be the (partial) layout under construction. The set \emph{unvisited} is the set of vertices of graph \emph{G} which are not in \emph{layout}. %The heuristic begins with initializing \emph{unvisited} to the vertex set\emph{V}(\emph{G}) in Step 1. In Step 2, a vertex $\emph{v}\in \emph{unvisited}$ of minimum degree is selected and is placed in the layout \emph{layout} in Step 3. In Step 6, a subset \emph{S} of \emph{layout} is identified which consists of those vertices of \emph{layout} which are adjacent to the least number of vertices of \emph{unvisited}. Now a subset \emph{P} of \emph{unvisited} is identified (Step 7) which consists of those vertices of \emph{unvisited} which are adjacent to the largest number of vertices of \emph{S}. Next a subset \emph{Q} of \emph{P} is determined which consists of those vertices of \emph{P} that are adjacent to the least number of vertices in unvisited. Now if there is more than one vertex in \emph{Q}, then a vertex $\emph{v}\in \emph{Q}$ is selected randomly and placed in the layout.

The time complexity of \emph{H}1 is $O(n^2\cdot \Delta(G))$.
%\begin{algorithm}[h]
%\begin{algorithm}

\begin{figure}[t]
\begin{description}
\setlength{\itemsep}{-2pt}
\item [Step 1:] $unvisited =\{1,\dots,n\}$
\item [Step 2:] Select a vertex $v\in unvisited$ such that $d_G(v)$ is least
\item [Step 3:] $layout = (v)$
\item [Step 4:] $unvisited = unvisited\setminus \{v\}$
\item [Step 5:] \textbf{while} $|layout| \neq n$
\item [Step 6:]	\hspace{0.5 cm}$S = \{v\in layout \colon v$ is adjacent to least number of vertices in $unvisited$\}
\item [Step 7:]	\hspace{0.5 cm}$P = \{w\in unvisited \colon w$ is adjacent to largest number of vertices in $S$\}
\item [Step 8:] 	\hspace{0.5 cm}$Q = \{u\in P \colon u$ is adjacent to least number of vertices in $unvisited$\}
\item [Step 9:] 	\hspace{0.5 cm}Select $v\in Q$ randomly
\item [Step 10:] \hspace{0.5 cm}$layout = (layout,v)$
\item [Step 11:] \hspace{0.5 cm}$unvisited = unvisited\setminus \{v\}$
\item [Step 12:] \textbf{endwhile}
\end{description}
\caption{Algorithm1: Heuristic \emph{H}1}
\end{figure}

%\end{alg}
%\end{algorithm}

%\emph{Time Complexity of H1}
%In Step 2, degree $\emph{d}_\emph{G}(\emph{v})$ of all the vertices is computed and then the vertex with least degree is identified. Therefore, Step 2 takes $\emph{O}(\emph{n}\cdot \bigtriangleup(\emph{G}))$ where $\bigtriangleup(\emph{G})$ denotes $ max_(\emph{v}\in \emph{V}(\emph{G}))⁡d_G (v)$. Step 4 can be computed in \emph{O}(\emph{n}) time. In Step 6, it is required to find adjacent vertices of $\emph{v}\in \emph{layout}$ in \emph{unvisited} which can be computed in $\emph{O}(\bigtriangleup(G))$ time. Since this is performed for all $ v\in layout$, therefore, in the worst case its time complexity is $\emph{O}(\emph{n}\cdot \bigtriangleup(\emph{G}))$. After this we need to find the subset of vertices of \emph{layout} which is adjacent to the least number of vertices of \emph{unvisited}. This can be computed in linear time. Therefore, time complexity of Step 6 is $\emph{O}(\emph{n}\cdot \bigtriangleup(\emph{G}))$. Similarly, time complexity of Step 7 and Step 8 is also $\emph{O}(\emph{n}\cdot \bigtriangleup(\emph{G}))$. Step 9 can be performed in \emph{O}(1) time and Step 10 and Step 11 can be completed in linear time. Hence, the time complexity of \emph{H}1 is $\emph{O}(\emph{n}^2\cdot \bigtriangleup(\emph{G}))$.
\subsection{Heuristic \emph{H}2}
Algorithm 2 presents heuristic \emph{H}2. %Heuristic \emph{H}2 begins by selecting a minimum degree vertex \emph{u} and adds it to the initially empty \emph{layout}. After this the vertices adjacent to \emph{u} are placed in the \emph{layout} in the ascending order of their degrees.% in Step 5. In Step 8, the set $degree = [degree_i]_i\in \{1,\dots,\emph{n}\}$, indexed by vertex identifiers, is constructed  where $degree_i$ for vertex \emph{i} is the number of adjacent vertices which are not in \emph{layout}. This set is used for selecting the next vertex and its neighbours for inclusion in Steps 9 to 16. This process continues till a complete layout is found.
 Time complexity of Heuristic \emph{H}2 is $O(n^2 \cdot \log n)$.

%\begin{algorithm}[h]
%\begin{alg}
\begin{figure}[t]
\begin{description}
\setlength{\itemsep}{-2pt}
\item [Step 1:] u = minimium degree vertex of a graph G
\item [Step 2:] $unvisited = V\setminus\{u\}$
\item [Step 3:] $layout = (u)$
\item [Step 4:] S = sort N(u) in ascending order of the degrees
\item [Step 5:] $layout = (layout,S)$
\item [Step 6:] $unvisited = unvisited\setminus S$
\item [Step 7:] \textbf{while} $|layout| \neq |V(G)|$
\item [Step 8:] \hspace{0.5 cm}$degree=\{degree_w=|N(w)\setminus layout|\colon w\in V(G)\}$
\item [Step 9:]	\hspace{0.5 cm}select a vertex v such that $degree_v$ is least but non-zero
\item [Step 10:]\hspace{0.5 cm}\textbf{if} $v \notin layout$
\item [Step 11:]\hspace{1 cm}$layout = (layout,v)$
\item [Step 12:]\hspace{1 cm}$unvisited = unvisited\setminus {v}$
\item [Step 13:]\hspace{0.5 cm}\textbf{endif}
\item [Step 14:]\hspace{0.5 cm}$S = N(v)\setminus layout$
\item [Step 15:]\hspace{0.5 cm}$S'$ = sort $S$ according to their values in degree in ascending order
\item [Step 16:]\hspace{0.5 cm}$layout = (layout,S')$
\item [Step 17:]\hspace{0.5 cm}$unvisited = unvisited\setminus S'$
\end{description}
\caption{Algorithm2: Heuristic \emph{H}2}
\end{figure}
%Step 18:\textbf{endwhile}
%\end{alg}
%\end{algorithm}

%\emph{Time Complexity of H2} \\
%In Step 1, minimum degree vertex of graph \emph{G} can be computed in $O(n\cdot \bigtriangleup(G))$ time. In worst case, Step 4 takes $O(n\cdot log n)$ time. In Step 8, \emph{$degree_w$} can be computed in $O(\bigtriangleup(G))$ time and the construction of set degree takes $O(n\cdot \bigtriangleup(G))$ time. Step 9, 12 and 17 can be completed in linear time in number of vertices. In worst case, time complexity of Step 15 is $O(n\cdot log n)$ . Hence, the time complexity of Heuristic \emph{H}2 is $O(n^2 \cdot log n)$ .
\subsection{Heuristic \emph{H}3}
This Heuristic is similar to \emph{H}2, but as opposed to the selection in Step 9 of Heuristic \emph{H}2, preference is given to a vertex \emph{v} in the partially constructed layout with least $\varphi(\emph{v})$. Time complexity of \emph{H}3 is same as \emph{H}2.
\section{Computational Experiments}
In this section we present the computational experiments that were performed to test the effectiveness of the proposed construction heuristics. The code was implemented in Matlab 7.0 and all the experiments were conducted on an Intel i3 based system with 4 GB RAM. The experiment to compare the construction heuristics was performed on the four sets of instances previously used for this problem \cite{noberto}. The test set includes Small, Harwell-Boeing (\emph{HB}) graphs, Grids and Trees. Since GVNS was tested on Harwell-Boeing (\emph{HB}) graphs, Grids and Trees in the literature, therefore, construction heuristic \emph{H}1 and GVNS are compared on these instances only. The order of graph ranges from 9 to 2916. For each heuristic \emph{H}1, \emph{H}2 and \emph{H}3, 30 runs were carried out and the best value of vertex separation was recorded. %A brief description of test instances is as follows:\\\\
%\emph{Small}: This set of instances consists of 84 graphs, whose optimal value is not known. In this data set the number of vertices and edges ranges from 16 to 24 and from 18 to 49, respectively.\\\\
%\emph{HB}: This set consists of 57 instances which is a subset of Harwell-Boeing Sparse Matrix Collection. The number of vertices and edges ranges from 24 to 960 and from 34 to 3721 respectively.\\\\
%\emph{Grids}: This set consists of 50 square grids. The vertex separation of $\emph{m}\times\emph{m}$ grid is \emph{m}.\\\\
%Trees: Let $\emph{T}(\lambda)$ be the set of trees with minimum number of nodes and vertex separation equal to $\lambda$. This set consists of 50 different trees: 15 trees in \emph{T}(3), 15 trees in \emph{T}(4) and 20 trees in \emph{T}(5). The number of vertices and edges ranges from 22 to 202 and from 21 to 201 respectively.
\subsection{Comparison between Construction Heuristics}
In this section the proposed construction heuristics are compared with those available in the literature. \emph{Random} represents randomly generated solution. \emph{C}1 and \emph{C}2 are the construction heuristics proposed in \cite{duarte} and \emph{HA}1, \emph{HA}2, \emph{HA}3, \emph{HA}4, \emph{HN}1, \emph{HN}2, \emph{HN}3 and \emph{HN}4 have been proposed in \cite{noberto}. Heuristics \emph{H}1, \emph{H}2 and \emph{H}3 have been presented in this paper. Table \ref{T1} presents the average vertex separation for different construction heuristics. The average vertex separation over all the instances are given in the last column $'Average'$. Results show that \emph{H}1 outperforms all the heuristics. Table \ref{T2} presents number of best solution achieved using each heuristic. In the table, the last column $'Sum'$ represents the total number of instances for which heuristic achieves the best value. Number in bold indicates that the value is best in that column. Heuristic \emph{H}1 achieves best result for 220 instances out of 248 instances which is equal to 88.71\% of the total instances. Similarly, \emph{H}2 and \emph{H}3 achieve best results for 43.55\% and 37.1\% instances of total instances. Among the existing heuristics, a maximum of 39.92\% of the total instances best solutions are obtained by \emph{HN}1.
\begin{table}
\caption{Average vertex separation for different construction heuristics}\label{T1}
  \centering
        \begin{tabular}{c c c c c c}
\hline
  \textbf{Heuristics} &\textbf{Grid(52)} &\textbf{Small(84)} &\textbf{Tree(50}) & \textbf{HB(62)} & \textbf{Average} \\
    \hline
 \textbf{Random} & 553.08 & 9.26 & 43.6 & 229.03 & 185.15 \\
  \textbf{\emph{C}1} & 29.5 & 6.18 & 4.66 & 40.1 & 19.24 \\
  \textbf{\emph{C}2} & \textbf{28.5} & 5.01 & 11.08 & 34.7 & 18.58 \\
  \textbf{\emph{HA}1} & 38.19 & 4.32 & 5.24 & 53.65 & 23.94 \\
  \textbf{\emph{HA}2} & 684.35 & 12.21 & 61.7 & 294.58 & 233.71 \\
  \textbf{\emph{HA}3} & 38.5 & 4.26 & 5.22 & 53.08 & 23.84 \\
  \textbf{\emph{HA}4} & 688 & 11.68 & 61.68 & 294.15 & 234.19 \\
  \textbf{\emph{HN}1} & 28.5 & 4.05 & 7.66 & 49.23 & 21.2 \\
  \textbf{\emph{HN}2} & 687.94 & 6.76 & 12.86 & 235.45 & 207.99 \\
  \textbf{\emph{HN}3} & 30.44 & 4.14 & 7.52 & 51.27 & 22.12 \\
  \textbf{\emph{HN}4} & 697.96 & 6.51 & 13.36 & 233.52 & 209.63 \\
  \textbf{\emph{H}1} & \textbf{28.5} & \textbf{3.29} & \textbf{4.1} & \textbf{30.53} & \textbf{15.6} \\
  \textbf{\emph{H}2} & 28.52 & 4.02 & 15.74 & 35.48 & 19.45 \\
  \textbf{\emph{H}3} & \textbf{28.5} & 4.28 & 11.52 & 37.05 & 19.07 \\
  \hline
  \end{tabular}
  \end{table}
\begin{table}
\caption{Number of best solutions achieved for different construction heuristics}\label{T2}
  \centering
\begin{tabular}{c c c c c c}
%\caption{Average vertex separation for different construction heuristics}
%\begin{center}
  \hline
  \textbf{Heuristics} &\textbf{Grid(52)} &\textbf{Small(84)} &\textbf{Tree(50}) & \textbf{HB(62)} & \textbf{Sum} \\
    \hline
 \textbf{Random} & 0 & 0 & 0 & 1 & 1 \\
  \textbf{\emph{C}1} & 0 & 1 & 12 & 1 & 14 \\
  \textbf{\emph{C}2} & \textbf{52} & 7 & 6 & 8 & 73 \\
  \textbf{\emph{HA}1} & 0 & 20 & 0 & 4 & 27 \\
  \textbf{\emph{HA}2} & 0 & 0 & 0 & 1 & 1 \\
  \textbf{\emph{HA}3} & 0 & 23 & 4 & 4 & 33 \\
  \textbf{\emph{HA}4} & 0 & 0 & 0 & 1 & 1 \\
  \textbf{\emph{HN}1} & \textbf{52} & 29 & 12 & 6 & 99 \\
  \textbf{\emph{HN}2} & 0 & 1 & 1 & 1 & 3 \\
  \textbf{\emph{HN}3} & 1 & 27 & 14 & 3 & 45 \\
  \textbf{\emph{HN}4} & 1 & 0 & 0 & 1 & 2 \\
  \textbf{\emph{H}1} & \textbf{52} & \textbf{80} & \textbf{50} & \textbf{38} & \textbf{220} \\
  \textbf{\emph{H}2} & 51 & 31 & 3 & 23 & 108 \\
  \textbf{\emph{H}3} & \textbf{52} & 25 & 1 & 14 & 92 \\
  \hline
\end{tabular}
\end{table}
\subsection{Comparison between \emph{H}1 and GVNS}
In this section, the best performing construction heuristic \emph{H}1 is compared with the metaheuristic General Variable Neighbouhood Strategy (GVNS) \cite{oro}. Average vertex separation obtained by GVNS and \emph{H}1 are listed in Table \ref{T3}. Numbers in bold are the minimum in that column. Results show that for Trees, \emph{H}1 performs better while performance of GVNS on HB instance is good. When the average is considered over all the instances, performance of GVNS and \emph{H}1 is comparable. Table \ref{T4} presents the number of best solutions achieved using each heuristic. For grids, GVNS and \emph{H}1 both are able to obtain optimal results. For trees, GVNS obtained optimal results for 40 instances out of 50 while \emph{H}1 obtained optimal results for all the tree instances. In case of HB instances, GVNS obtained best results for 44 instances out of 62 while \emph{H}1 was able to obtain optimal results for 33 instances only. When all the graphs are considered GVNS obtained best results in 136 cases out of 164 which is equal to 82.9\% of all input instances while \emph{H}1 obtained optimal results in 135 cases which is equal to 82.3\% of all input instances.
\begin{table}
\caption{Average vertex separation for GVNS \cite{oro} and \emph{H}1}\label{T3}
  \centering
\begin{tabular}{c c c c c}
\hline
 \textbf{Heuristics} &\textbf{Grid(52)} &\textbf{Tree(50}) & \textbf{HB(62)} & \textbf{Average}\\
    \hline
  \textbf{GVNS} & \textbf{28.5} & 4.3 & \textbf{24.6} & 19.85 \\
  \textbf{\emph{H}1} & \textbf{28.5} & \textbf{4.1} & 30.53 & 21.83 \\
  \hline
\end{tabular}
\end{table}
 \begin{table}
 \caption{Number of best solutions achieved using GVNS and \emph{H}1}\label{T4}
  \centering
\begin{tabular}{c c c c c}
 \hline
 \textbf{Heuristics} & \textbf{Grid(52)} & \textbf{Tree(50}) & \textbf{HB(62)} & \textbf{Sum} \\
    \hline
  \textbf{GVNS} & \textbf{52} & 40 & \textbf{44} & \textbf{136} \\
  \textbf{\emph{H}1} & \textbf{52} & \textbf{50} & 33 & 135 \\
  \hline
\end{tabular}
\end{table}
\section{Conclusion}
In this paper, we have proposed three polynomial time construction heuristics and compared with other construction heuristics. It was observed that these construction heuristics outperform other construction heuristics given in literature. The best performing construction heuristic \emph{H}1 was also compared with the state-of-the-art metaheuristic GVNS. This construction heuristic achieves best results in larger number of cases than GVNS. Since a good initial population is expected to converge faster to a near optimal/optimal solution, therefore, initial solutions generated using this construction heuristic in any metaheuristic will lead to high quality solutions.

\end{document}